\documentclass[12pt]{iopart}
\usepackage{iopams}
\begin{document}

\title [Two paths towards circulation time derivative] {Two paths towards circulation time derivative (Maxwell's $\mathfrak E$ revisited)}
\author{D V Red\v zi\' c}

\address{$^1$Faculty of Physics, University of Belgrade, PO
Box 44, 11000 Beograd, Serbia} \eads{\mailto{redzic@ff.bg.ac.rs}}

\begin{abstract}
The time derivative of the circulation of a vector field $\bi A$
over a moving and deforming closed curve, $\frac {\rmd}{\rmd t}\oint
\bi A \cdot \rmd \bi r$, is computed in two ways, with and without
bringing the time derivative under the integral sign. As a
by-product, the computations reveal that the conceptualization of
Faraday's law of electromagnetic induction may depend on which of
the two methods is employed. The discussion presented provides an
unexpected argument in favor of Maxwell's mysterious choice for his
electromotive intensity $\mathfrak E$, made in Article 598 of his
{\it Treatise}.
\end{abstract}
%\pacs{XXX}
%\maketitle
\section{Introduction}
Recently, we expounded how Maxwell had arrived, through an ingenious
analysis of Faraday's law of electromagnetic induction given in
Article 598 of his {\it Treatise} \cite {M54}, at a general
expression for his electromotive intensity $\mathfrak E$ in a moving
medium:
\begin{equation}
\mathfrak E  = \bi v \times \bi B - \frac {\partial \bi A}{\partial
t}  - \bnabla \Psi \, ;
\end{equation}
here $\bi v$ is the velocity of an infinitesimal portion
(`particle') of the medium, $\bi B  =  \bnabla \times \bi A$ is the
magnetic flux density, $\bi A$ is the vector potential, and a scalar
field $\Psi$ is Maxwell's {\it electric potential} \cite{DR1}. We
recalled that various authors claimed that Maxwell should have
included a term $-\bnabla(\bi A \cdot \bi v)$ in expression (1), as
is strongly suggested by his derivation of Article 598. Namely, in
Maxwell's computation of the negative time derivative of the
circulation of $\bi A$, $-\frac {\rmd}{\rmd t}\oint \bi A \cdot \rmd
\bi r$, two terms of his result are expressed through $-\oint d(\bi
A \cdot \bi v) = -\oint \bnabla(\bi A \cdot \bi v)\cdot \rmd \bi r$.
However, Maxwell mysteriously leaves out the gradient term
$-\bnabla(\bi A \cdot \bi v)$ in his final version of the integrand,
noting simply that it vanishes when integrated round a closed curve,
and introduces a brand-new term $-\bnabla \Psi$, `for the sake of
giving generality' to the expression (1) for $\mathfrak E$. The
situation is even more curious, taking into account that the
alternative expression for the electromotive intensity,

\begin{equation}
\mathfrak E_{\mbox {\scriptsize {HWT}}}  = \bi v \times \bi B -
\frac {\partial \bi A}{\partial t}  - \bnabla \Psi  - \bnabla (\bi A
\cdot \bi v)\, ,
\end{equation}
as proposed by Helmholtz \cite{HH}, Watson \cite{W}, and J J Thomson
\cite{M54} (vol 2, p 260), see also \cite {JJT, JB}, complies
perfectly with Maxwell's {\it general} principle of relativity
applied to the Faraday's law, as is demonstrated in \cite{DR2}.

Perhaps surprisingly, it turns out that the appearance of the term
$\bnabla(\bi A \cdot \bi v)$ is an artefact of the specific path
employed by Maxwell for computing $\frac {\rmd}{\rmd t}\oint \bi A
\cdot \rmd \bi r$. Namely, in an alternative computation path, the
controversial term simply does not appear. This fact which seems to
be little known, unfortunately, had escaped our attention during the
writing of \cite{DR1,DR2}, while it was implicit in \cite{DR3}.

In the present note, we first outline Maxwell's computation of
$\frac {\rmd}{\rmd t}\oint \bi A \cdot \rmd \bi r$ over a moving and
deforming closed curve, given in Article 598, which involves the
non-obvious step of bringing the time derivative under the integral
sign. Then we give the alternative, simpler computation of $\frac
{\rmd}{\rmd t}\oint \bi A \cdot \rmd \bi r$, applying the
Kelvin-Stokes theorem twice, which avoids bringing the time
derivative under the integral sign, and which is free from the term
$\bnabla(\bi A \cdot \bi v)$. Both computations could be useful from
didactic point of view. Also, it could be inspiring for the student
to learn that the conceptualization of Faraday's law may depend on
the specific path chosen for computing $\frac {\rmd}{\rmd t}\oint
\bi A \cdot \rmd \bi r$. Moreover, the simpler computation appears
to provide an unexpected vindication of Maxwell's happy and
controversial choice for $\mathfrak E$, one of the key concepts of
his electromagnetic theory and the progenitor of the Lorentz force
expression.

\section {Two paths for the computation of circulation time derivative}

\subsection {Maxwell's path}

For the sake of completeness, we outline Maxwell's computation of
the total time derivative of the circulation of an arbitrary,
continuous and differentiable vector field $\bi A(\bi r,t)$ over a
moving and deforming closed curve $C(t)$ at the instant $t$.
Contrary to Maxwell, who writes everything in the Cartesian form, we
employ the modern vector notation, benefiting from Hamilton's
operator $\bnabla$, keeping, however, the spirit of Maxwell's
argument.\footnote [1] {Maxwell's original argument, free from
$\bnabla$, is presented in full detail in \cite{DR1}, and also,
almost literally, and in a somewhat complemented form, in
\cite{AY}.}

Maxwell writes the circulation as, in modern notation,

\begin {equation}
\oint_{C(t)}\bi A \cdot \rmd \bi r = \int_0^{s_{max}(t)}\left
(A_x\frac{\partial x}{\partial s} + A_y\frac{\partial y}{\partial s}
+ A_z\frac{\partial z}{\partial s}\right )\rmd s\, ,
\end{equation}
where $\bi r = \bi r(s,t)$ is the position vector of a point of the
contour, parameter $s$ is the arc length of the point considered
{\it at the instant} $t$, and $s_{max}(t)$ is the total length of
the contour at that instant. Since the circulation of $\bi A$ refers
to the fixed $t$, $\rmd \bi r$ is the partial differential of $\bi
r$ with respect to $s$, that is $\rmd \bi r = \frac {\partial \bi
r}{\partial s}\rmd s \equiv \rmd_s\bi r$. Note that Maxwell takes
tacitly that the parametrization which refers to the {\it fixed}
instant $t$ suffices for describing the moving and deforming contour
also in subsequent instants so that $s$ is time-independent. (This
is of course correct; as can be seen, the fact that the total length
of the moving contour is time-dependent is irrelevant, there is a
bijection between the corresponding two sets of points.) Thus, a
point $\bi r(s,t)$ at the instant $t + \rmd t$ becomes $\bi
r(s,t+\rmd t) = \bi r(s,t) + \bi v(s,t)\rmd t$ where $\bi v(s,t) =
\frac{\partial \bi r(s,t)}{\partial t}$ is the instantaneous
velocity of the point relative to the Cartesian coordinate system
chosen.

To compute the time derivative of the circulation in the case of a
moving and deforming contour $C(t)$, Maxwell takes the time
derivative inside the integral sign\footnote [2] {The validity of
this step is not very obvious and a proof is given in the appendix
of \cite{DR1}, arriving at equation (7) directly from the definition
of $\frac {\rmd}{\rmd t}\oint_{C(t)}\bi A \cdot \rmd \bi r$. An
alternative proof, involving a renormalization of the variable $s$
at each instant $t$ is presented in \cite{AY}. While the
renormalization procedure is mathematically expedient, it is not
indispensable, the parametrization at {\it one} instant suffices, cf
the appendix of \cite{DR1}. As can be seen, another way of
vindicating this step would be to invoke the Leibniz rule for
differentiating an integral function, cf, e.g., \cite{EB}, taking
into account that $s$ is time-independent.} and thus
\begin {equation}
\frac {\rmd}{\rmd t}\oint_{C(t)}\bi A \cdot \rmd \bi r =
\oint_{C(t)}\frac {\rmd}{\rmd t}(\bi A \cdot \rmd \bi r) =
\oint_{C(t)}\left (\frac {\rmd \bi A}{\rmd t} \cdot \rmd \bi r + \bi
A \cdot \frac {\rmd}{\rmd t}\rmd \bi r\right )\, ,
\end{equation}
The differentiations yield

\begin {equation}
\frac {\rmd \bi A}{\rmd t} = \frac {\partial \bi A}{\partial t} +
(\bi v \cdot \bnabla)\bi A\, ,
\end{equation}
and

\begin {equation}
\frac {\rmd}{\rmd t}\rmd \bi r = \rmd \bi v\, ,
\end{equation}
where $\rmd \bi v = \frac {\partial \bi v}{\partial s}\rmd s \equiv
\rmd_s\bi v$, since $s$ is time-independent. Maxwell thus obtains

\begin {equation}
\frac {\rmd}{\rmd t}\oint_{C(t)}\bi A \cdot \rmd \bi r =
\oint_{C(t)}\frac {\partial \bi A}{\partial t} \cdot \rmd \bi r +
\oint_{C(t)}[(\bi v \cdot \bnabla)\bi A] \cdot \rmd \bi r +
\oint_{C(t)}\bi A \cdot \rmd \bi v\, .
\end{equation}
Equation (7) is, basically, Maxwell's equation (2) of Art. 598,
written in compact form, employing the modern vector
notation.\footnote [3] {Note that our expression $\frac {\rmd}{\rmd
t}\rmd \bi r$ is nothing but Maxwell's $\left [\frac {\rmd}{\rmd
t}\left (\frac {\partial \bi r}{\partial s}\right ) \right ] \rmd s$
(clearly implicit in Art. 598), since $s$ is time-independent.}

Now express $(\bi v \cdot \bnabla)\bi A$ via the well-known vector
identity

\begin {equation}
\bi v \times (\bnabla \times \bi A) = \bnabla(\bi v\cdot \bi A^*) -
(\bi v \cdot \bnabla)\bi A \, ,
\end{equation}
where the asterisk in the expression $\bnabla(\bi v\cdot \bi A^*)$
indicates that $\bnabla$ operates only on $\bi A$. Employing also
equation

\begin {equation}
\bnabla(\bi v\cdot \bi A^*)\cdot \rmd \bi r = \rmd (\bi v\cdot \bi
A^*) = \bi v\cdot \rmd \bi A\, ,
\end{equation}
one obtains

\begin {equation}
[(\bi v \cdot \bnabla)\bi A] \cdot \rmd \bi r = [(\bnabla \times \bi
A) \times \bi v]\cdot \rmd \bi r + \bi v \cdot \rmd \bi A\, ,
\end{equation}
Inserting (10) into (7) yields

\begin {equation}
\frac {\rmd}{\rmd t}\oint_{C(t)}\bi A \cdot \rmd \bi r =
\oint_{C(t)}\left [\frac {\partial \bi A}{\partial t} + (\bnabla
\times \bi A) \times \bi v\right ]\cdot \rmd \bi r  +
\oint_{C(t)}\rmd (\bi A \cdot \bi v)\, .
\end{equation}
or, equivalently,

\begin {equation}
\frac {\rmd}{\rmd t}\oint_{C(t)}\bi A \cdot \rmd \bi r =
\oint_{C(t)}\left [\frac {\partial \bi A}{\partial t} + (\bnabla
\times \bi A) \times \bi v + \bnabla (\bi A \cdot \bi v)\right
]\cdot \rmd \bi r \, .
\end{equation}
Finally, noting that the last integral in eq. (11) vanishes since it
is taken round the closed curve, Maxwell arrives at

\begin {equation}
\frac {\rmd}{\rmd t}\oint_{C(t)}\bi A \cdot \rmd \bi r =
\oint_{C(t)}\left [\frac {\partial \bi A}{\partial t} + (\bnabla
\times \bi A) \times \bi v\right ]\cdot \rmd \bi r \, .
\end{equation}

Equation (13) is a purely mathematical and general result valid for
an arbitrary moving and deforming closed curve $C(t)$ that remains
continuous and closed during its motion, and for arbitrary,
continuous and differentiable vector field $\bi A(\bi r, t)$ and
velocity field $\bi v(\bi r, t)$. Note that the appearance of the
controversial term $\bnabla (\bi A \cdot \bi v)$ in eq. (12) is a
consequence of computing $\frac {\rmd}{\rmd t}(\bi A \cdot \rmd \bi
r)$.

\subsection {The simpler path}

Now we present a simpler computation of $\frac {\rmd}{\rmd
t}\oint_{C(t)}\bi A \cdot \rmd \bi r$, which avoids bringing the
time derivative under the integral sign, and avoids (explicit)
parametrization of the curve $C(t)$.

The time derivative of the circulation of $\bi A$ is by definition:

\begin {equation}
\frac {\rmd}{\rmd t}\oint_{C(t)}\bi A (\bi r, t) \cdot \rmd \bi r =
\frac {\oint_{C(t + \rmd t)}\bi A (\bi r, t + \rmd t)\cdot \rmd \bi
r - \oint_{C(t)}\bi A (\bi r, t)\cdot \rmd \bi r}{\rmd t}\, .
\end{equation}

A Taylor series expansion in the first integral on the right hand
side of eq. (14) yields

\begin {equation}
\fl \oint_{C(t + \rmd t)}\bi A (\bi r, t + \rmd t)\cdot \rmd \bi r =
\oint_{C(t + \rmd t)}\bi A (\bi r, t)\cdot \rmd \bi r + \oint_{C(t +
\rmd t)}\frac {\partial \bi A (\bi r, t)}{\partial t} \rmd t \cdot
\rmd \bi r\, ,
\end{equation}
and applying the Kelvin-Stokes theorem to the second integral on the
right hand side of eq. (14) one has

\begin {equation}
\oint_{C(t)}\bi A (\bi r, t)\cdot \rmd \bi r =
\int_{S[C(t)]}[\bnabla \times \bi A (\bi r, t)]\cdot \rmd \bi S\, ,
\end{equation}
where $S[C(t)]$ is any open surface bounded by the closed curve
$C(t)$. Choosing for $S[C(t)]$ a surface which consists of a ribbon
swept by the moving contour during the time interval $\rmd t$ and a
surface $S[C(t + \rmd t)]$ (any open surface bounded by the closed
curve $C(t + \rmd t)$), the surface integral becomes

\begin {equation}
\fl \int_{S[C(t)]}[\bnabla \times \bi A (\bi r, t)]\cdot \rmd \bi S
= \oint_{C(t)}[\bnabla \times \bi A (\bi r, t)]\cdot (\rmd \bi r
\times \bi v \rmd t) + \int_{S[C(t + \rmd t)]}[\bnabla \times \bi A
(\bi r, t)]\cdot \rmd \bi S\, ,
\end{equation}
where $\bi v$ is the instantaneous velocity of the circuit element
$\rmd \bi r$ at the instant $t$. Transforming the right hand side of
eq. (17), rearranging terms in the first integral through a cyclic
permutation and applying the Kelvin-Stokes theorem to the second
integral, one obtains

\begin {equation}
\fl \oint_{C(t)}\bi A (\bi r, t)\cdot \rmd \bi r = \oint_{C(t)}\rmd
\bi r \cdot \{\bi v \rmd t \times [\bnabla \times \bi A (\bi r,
t)]\} + \oint_{C(t + \rmd t)}\bi A (\bi r, t)\cdot \rmd \bi r \, .
\end{equation}

Finally, inserting expressions (15) and (18) into the right-hand
side of eq. (14), taking into account that

\begin {equation}
\lim_{\rmd t \rightarrow 0}\oint_{C(t + \rmd t)}\frac {\partial \bi
A (\bi r, t)}{\partial t} \cdot \rmd \bi r = \oint_{C(t)}\frac
{\partial \bi A (\bi r, t)}{\partial t} \cdot \rmd \bi r\, ,
\end{equation}
the result (13) follows.

\section {Concluding comments}

The above discussion reveals that the conceptualization of Faraday's
induction law may depend on the specific path employed for computing
$\frac {\rmd}{\rmd t}\oint \bi A \cdot \rmd \bi r$. The simpler
computation path, applying the Kelvin-Stokes theorem twice, which
avoids bringing the time derivative under the integral sign, does
not yield the controversial term $\bnabla(\bi A \cdot \bi v)$. Thus
the issue of its inclusion into Maxwell's original expression for
the electromotive intensity $\mathfrak E$ is basically a
pseudo-problem. Namely, it seems reasonable to take that a quantity
whose appearance depends on the specific path chosen for computing
the {\it physical} quantity, $\frac {\rmd}{\rmd t}\oint \bi A \cdot
\rmd \bi r$, may have but a spurious physical meaning. Consequently,
the present note provides an unexpected argument in favor of
Maxwell's mysterious choice for $\mathfrak E$.

\section*{Acknowledgment}
My work is supported by the Ministry of Science and Education of the
Republic of Serbia, project No. 171028.

\end{document}